# Novel Improvement for Nonlinear Compatibility of Least Mean Square Adaptive Algorithm

Zhengyang Zhang[1]

[1]University of Cambridge

**Abstract**

In order to improve the least mean squares (LMS) adaptation algorithm to accommodate the nonlinear transfer function, and to adjust the coefficients of adaptive filter during the actual implement of bias voltage and signal amplitude, methods are proposed and simulated to develop a nonlinear adaptive filter. The inputs to LMS are replaced by the derivatives of traditional inputs, and the step for each training iteration is adaptively controlled by the difference between target signal and actual signal. The simulation utilizes the implementation of Nyquist pulses optical sampling and works as a digital signal processing pre-compensation to reduce influence of the frequency responses on wires and devices. The simulation result shows promising improvement with the modified adaptation algorithm method in tackling Mach Zehnder modulator's non-monotonic transfer function.

## 1. Introduction

The least mean squares (LMS) adaptation algorithm is one of the most common methods for gradian decent in training process of adaptive equalizers and filters [1] . The LMS algorithm is an effective and efficient algorithm to optimize the equalizer coefficients by minimizing the mean square error (MSE)

[11] . The algorithm updates the tap weights of a feedforward equalizer by formula [17] :

$$c_i \leftarrow c_i - \mu_s(y-z)x_i \tag{1}$$

where $y-z$ is the error introduced earlier and $x_i$ is the signal from the $i$th delay-line tap feeding weight $c_i$. The square error is $(y-z)^2$, and thus, its gradient with respect to the weights, $c_i$, follows as $2(y-z)x_i$ [13] . The Eq. (1) follows this negative gradient in every adaptation step, the size of which is controlled by $\mu_s$. It is on average equivalent to taking one larger step following the gradient of the MSE to take many steps controlled by $\mu_s$ following the gradient of the square error [14] . The least mean square is equivalent to performing a stochastic gradient descent on mean square error surface [15] .

The LMS can reach best performance when it is utilized to equalize the input signal or pre-compensate signal that has linear relation with the input signal to LMS. However, when the transfer function between the input signal to LMS and the signal pre-compensated by LMS is non-linear, the effectiveness of LMS reduces. What's worse, the LMS is not able to equalize the signal correctly, when the transfer function is non-monotonic, since the non-monotonic transfer function changes the heights of low points and high points in gradian map.

Because of aforementioned reason, implementation of LMS is limited, as most engineering applications have nonlinear transfer function. Non-monotonic transfer functions, such as raise cosine function in Mach Zehnder modulator and Lorenz function, also exist in a number of devices and applications [2] . To tackle the disadvantage of LMS, a novel improvement for LMS is proposed and simulated in the paper.

## 2. Principle

Considering the disadvantage of conventional LMS, a novel improvement of least mean square algorithm is proposed to expand application field to include implementations with non-monotonic transfer function.

## 2.1 Derivatives Utilized as Inputs of Improved Least Mean Squares (LMS) Adaptive Algorithm

The first part of innovative solution is to employ the derivatives of signals as the inputs of least mean square module, which will not heavily increase the complexity of the system in simulation.

There are mainly two reasons for the solution. First, the derivative has the information of signal's changing trend, since the integral of differential of function equals the original function plus an unsure constant.

Second, the derivative of even function is always odd function. Most non-monotonic functions used in engineering applications are even functions, such as raise cosine function and Lorenz function. The derivatives can offer a monotonic region. Taking the Mach Zehnder modulator (MZM) as an example, its transfer function is raise cosine function. The derivative of the transfer function of Mach Zehnder modulator is monotonic and not symmetric in operating region for pulse compression. Figure 1 shows that the original transfer function has different monotonicity in region $(\frac{V_\pi}{2}, V_\pi)$ and region $(V_\pi, \frac{3}{2}V_\pi)$, while the derivative of the transfer function has consistent monotonicity in operating region $(\frac{V_\pi}{2}, \frac{3}{2}V_\pi)$. Figure 1 shows the differential of Mach Zehnder modulator transfer function in compared with original Mach Zehnder modulator transfer function.

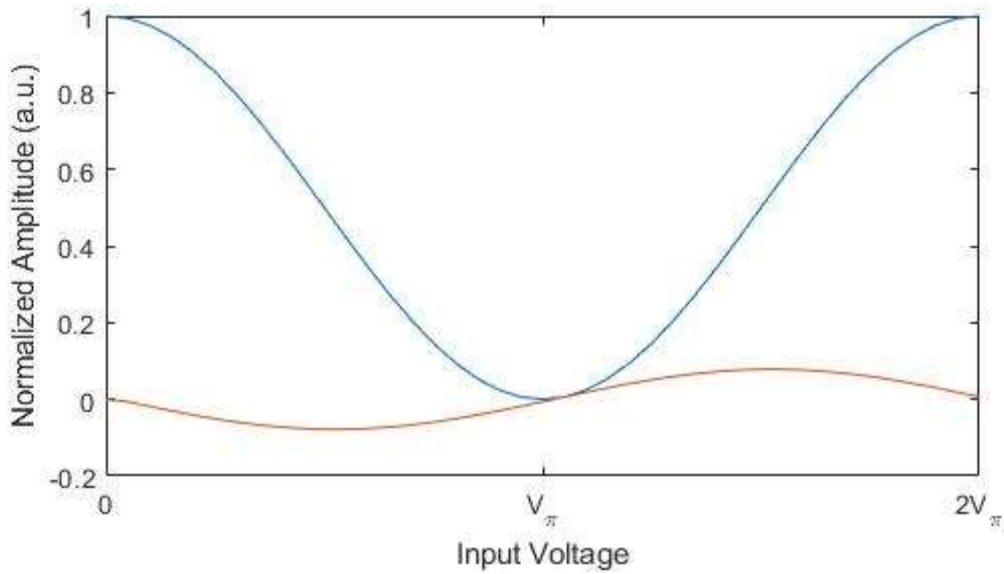

Figure 1. transfer function (blue) and differential of transfer function (orange)

## 2.2. Adaptive Adaptation Step

According to Eq. (1), the speed of gradient descent is controlled by the adaptation step, $\mu_s$ [17] . In the paper, the adaptation step is defined by the difference between the ideal output of Mach Zehnder modulator and the actual output of Mach Zehnder modulator, which is refreshed after each train iteration.

The adaptive adaptation step compensates for the loss of constant component when using derivatives of signals as input. Though the method of utilizing derivatives can solve the challenge of non-monotonic transfer function, the method also loss the information of constant value due to the nature of derivatives. Without the further assistant of adaptive adaptation step, the compensate signal can show a great increase of constant value in simulation. The adaptive adaptation step implemented in the paper is different from the common method in signal processing field, which utilize the absolute value of the difference of ideal signal and actual signal, and disregard the negative and positive symbol of the difference. In the paper, the adaptive adaptation step uses true value of the difference, offering the LMS adaptive algorism information about compensating direction blurred by the non-monotonic transfer function.

The speed of gradient descent is also benefited by the adaptation step. At the start of training process, the difference is larger. When the adaptive filter provides the best pre-compensation result, the difference automatically decreases, and the adaptation step becomes smaller preventing the taps oscillating instead of converging.

## 3. Simulation Schematics

The simulation takes the background of an actual engineering problem, which utilizes MZM as pulse compressor to achieve narrow peak Nyquist pulse train [2] . The work reported in [2] was forced to use linear adaptive filter due to the difficulty that the LMS is not compatible with non-monotonic transfer function. The simulation in this paper shows how the challenge is tackled by implementing the improved LMS algorism. The signal flow is illustrated in Figure 2.

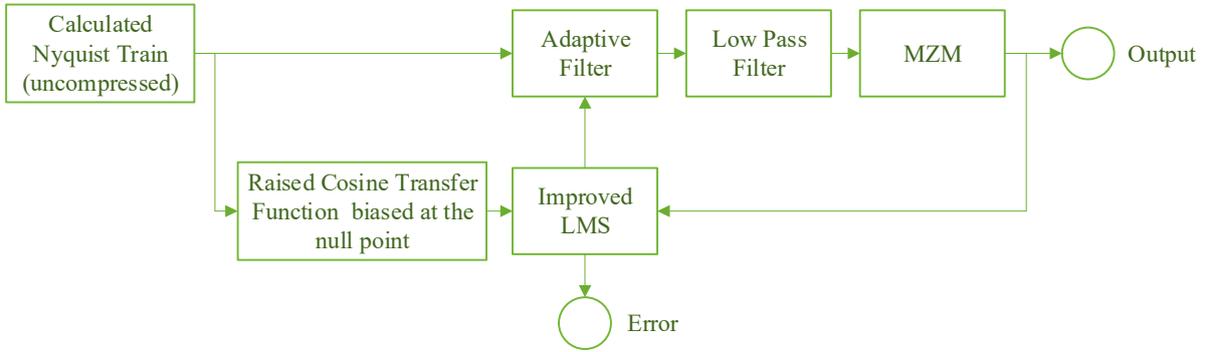

Figure 2. main processes involved in the project

The preparation of simulation included coding single Nyquist pulse train. The Nyquist pulse is firstly generated according to the Eq. (2) with sampling frequency at 10 THz [2] .

$$h(t) = \frac{\sin\left(\frac{\pi t}{T_s}\right)\cos\left(\frac{\pi \beta t}{T_s}\right)}{\left(\frac{\pi t}{T_s}\right)\left(1-\left(\frac{2\beta t}{T_s}\right)^2\right)} \quad (2)$$

where $T_s$ is the pulse duration between zero crossings and $\beta$ is the roll-off factor.

In the project, the Nyquist pulse is simulated in MatLab. The single Nyquist pulses with different roll-off factor, $\beta$, can then be simulated in time domain. The $T_s = 25ps$ and $\beta = 0.5$ are chosen for the simulation in project.

Having generated single Nyquist pulse, the Dirac comb is then employed to convolve with single Nyquist pulse to generate Nyquist pulses train. The process can be expressed by functions [12] :

$$h_T(t) = \text{Ш}_T(t) * h(t) \tag{3}$$

$$\text{Ш}_T(t) \triangleq \sum_{k=-\infty}^{\infty} \delta(t - kT) = \frac{1}{T}\text{Ш}\left(\frac{t}{T}\right) \tag{4}$$

for given period $T$, where $\delta(t)$ is the Dirac delta function. In the project, the pulse repetition rate is set to 10 GHz. The Nyquist pulses train is plotted in time domain in Figure 3 below.

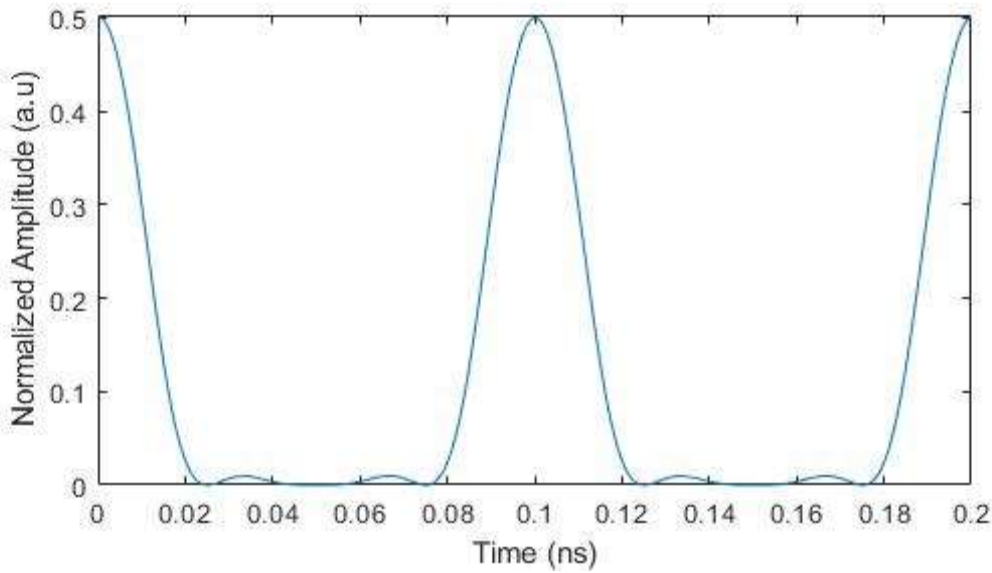

Figure 3. Nyquist pulses train in time domain

The simulation then came to the effect of cables and amplifiers' frequency response. The frequency bandwidths of single Nyquist pulse and Nyquist pulses train are calculated as comparison. The effect of cables and amplifiers' frequency response is simulated in this project by a low pass filter. The low pass filter has properties such as passband frequency at 1 Hz, stopband frequency at 5 GHz, stopband attenuation at 80dB, density factor at 20, and other parameters which is adaptively designed by MatLab FilterDesigner. The frequency response of the low pass filter is simulated in Figure 4.

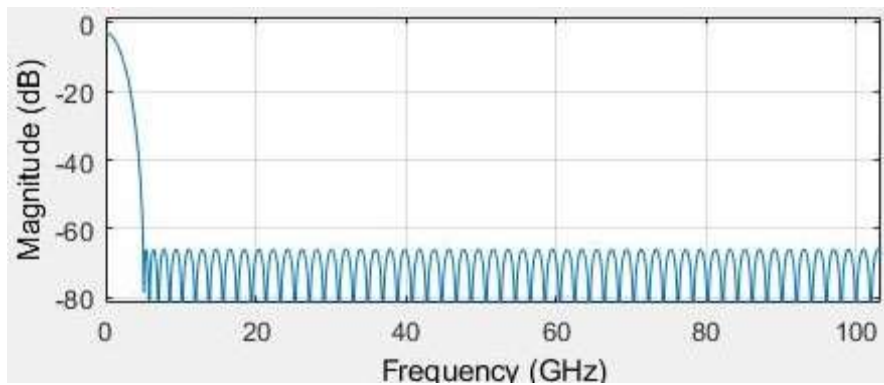

Figure 4. frequency response of the low pass filter

Finally, the power transfer function of MZM is used to simulate the behaviour of MZM [2] [15] [16] .

## 4. Simulation Result

The adaptive adaptation step is proposed in this project. The filter can also reach a stable position after gradient descent. The simulation result is shown in Figure 5

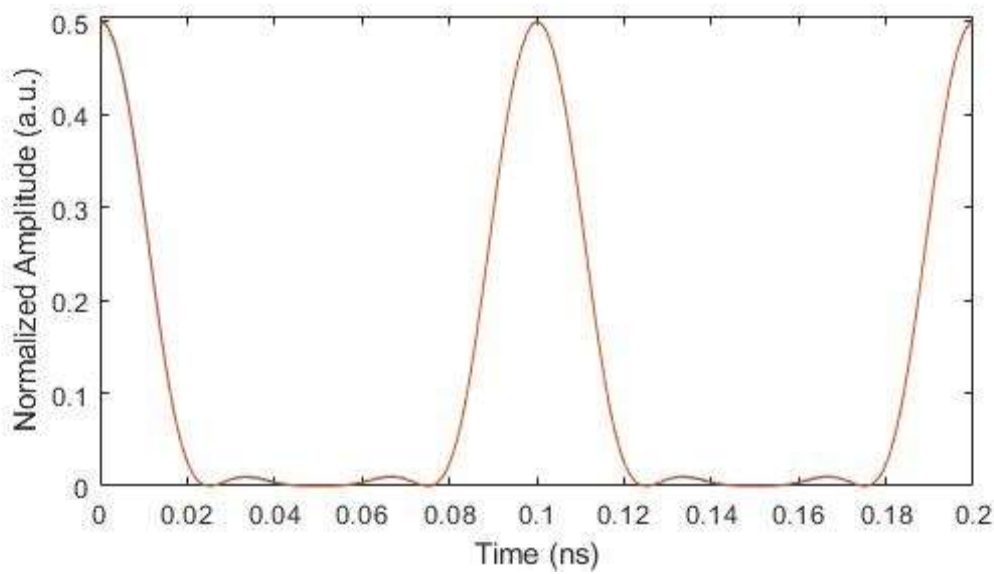

Figure 5. ideal MZM output signal (blue) and actual signal after compensation (orange)

To further identify how identical these two signals are, the difference of actual compensated signal and ideal signal is plotted in

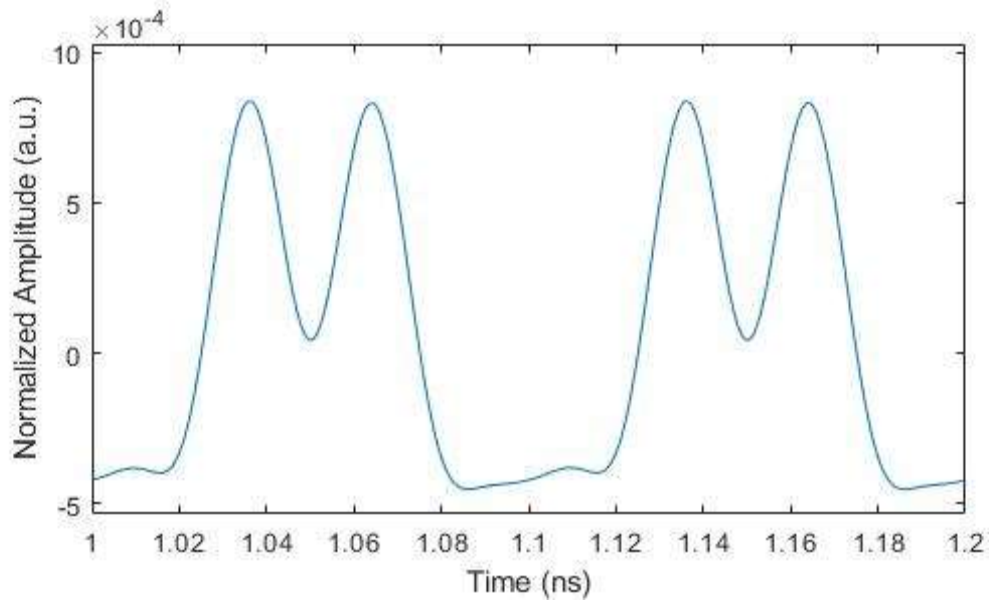

Figure 6. difference of actual compensated signal and ideal signal

The amplitude of difference is 1000 times smaller than the signals. Thus, the two signals can be seen as identical from engineering prospective.

## 5. Discussions

### 5.1. Comparison with Conventional LMS

With conventional LMS, the trail signal after hundreds of iterations has little correlation peak with the ideal signal, which forces the training iterations to stop.

To further investigate the reason, the trial signals in each training iteration is plotted and inspected, inside which the trial signals from 100$^{th}$ iteration to 106$^{th}$ iteration show irregular behavior.

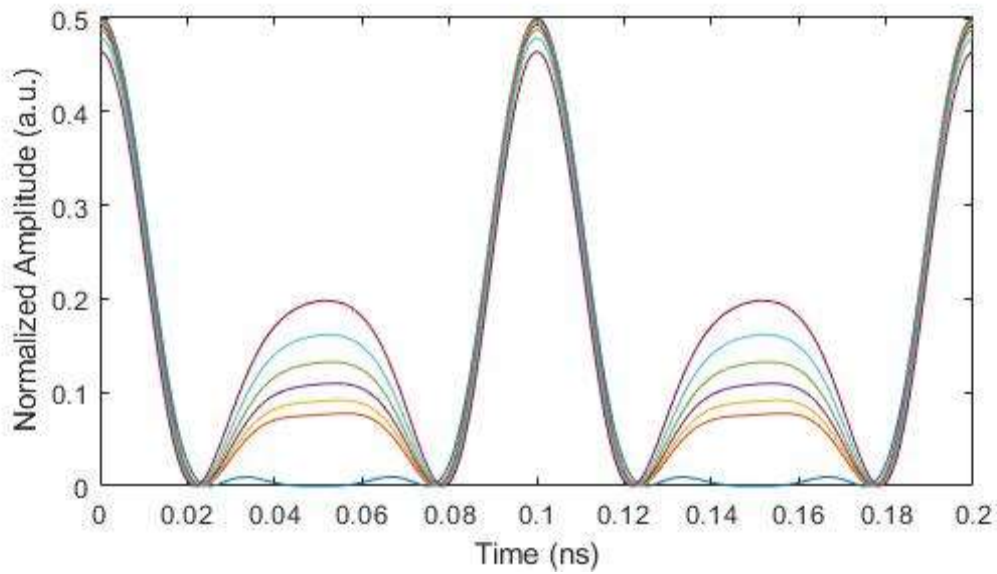

Figure 7. target signal (blue) and trial signals in 100$^{th}$ iteration (orange), 101$^{st}$ iteration (yellow), 102$^{nd}$ iteration (purple), 103$^{rd}$ iteration (green), 104$^{th}$ iteration (lighter blue), and 105$^{th}$ iteration (brown)

Since the purpose of utilizing least mean square algorithm is to reduce the difference and error between target waveform and pre-compensated waveform, the newer signal in Figure 7 should be closer to the target signal in comparison with the older signal. When observing curves between 30 ps and 70 ps, the simulation results, on the contrary, show the 100$^{th}$ iteration signal is at the bottom, while the 105$^{th}$ iteration signal is at the top. The error between target signal and trial signal increases after iterations of training by least mean square algorithm which is intended to decrease the error.

To investigate the reason of opposite modification direction, the input of the Mach Zehnder modulator is plotted. If the input signal is identical with the ideal Nyquist pulses train signal, the output should be identical to the target waveform.

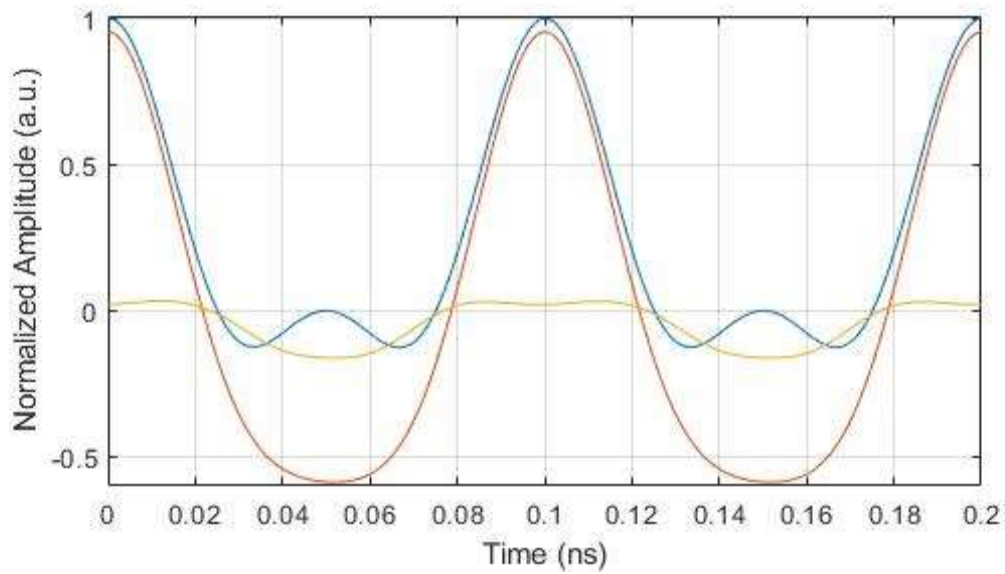

Figure 8. the ideal Nyquist pulses train signal (blue), the actual input signal (orange), and error calculated by LMS (yellow)

Because the principle of least mean square algorithm is to gradient decent to the least error, the symbol of error defines the direction of gradient decent. Since the final adaptive filter pre-compensated the input of the Mach Zehnder modulator, instead of output. The least mean square algorithm should direct the actual signal to increase at the time from 30 ps to 70 ps, and to increase at the time from 90 ps to 110 ps, in order to pre-compensate the actual input signal to be identical with the ideal Nyquist pulses train signal. However, when looking at the error curve in Figure 8, the error has different symbol at two time slots, leading the actual signal to different directions at the two time shots, which is opposite to previous inference.

One possible explanation is symmetrical property of Mach Zehnder modulator transfer function. As seen from Figure 8, the wrong symbol of error occurs at the negative parts of ideal Nyquist pulses train signal and the actual input signal, where the error should be a positive number, while on the contrary, the error calculated by comparing outputs of Mach Zehnder modulator as the inputs of least mean square algorithm is negative number. The symmetrical property of Mach Zehnder modulator, will convert every negative input into positive output, flipping up the negative part of signal to positive, when Mach Zehnder modulator is bias at $V_\pi$ operating as a pulse compressor.

**5.2. Generality of the Improved Method**

Another non-monotonic transfer function commonly implemented is the Lorenz. The simulation is run with similar settings, but the transfer function is replaced by Lorenz function.

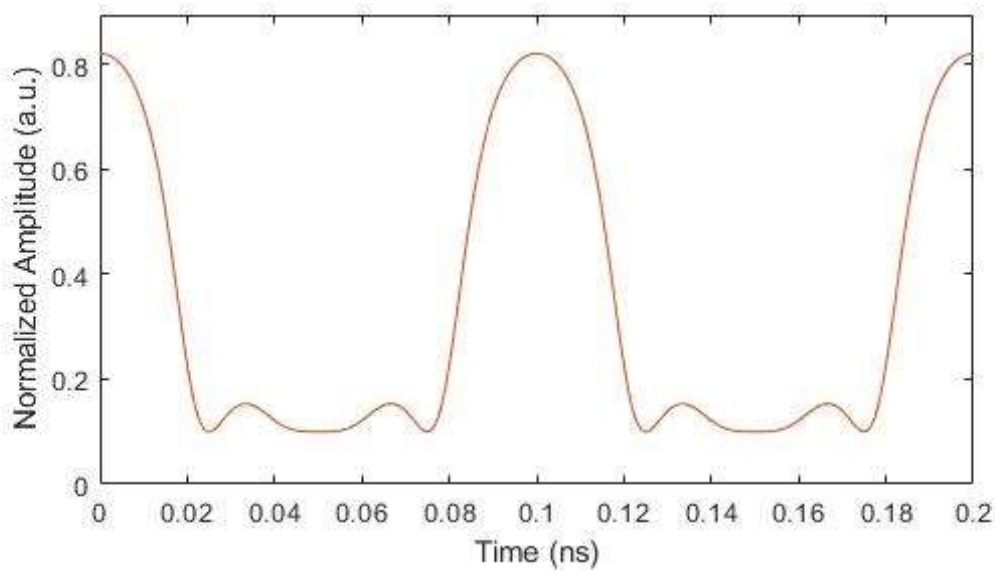

Figure 9. Improved LMS performance with Lorenz transfer function

Figure 9 shows the improved method can also be implemented when the transfer function is Lorenz. It further proves the generality of the improved method, and proves the analysis of the reason why LMS has difficulty in solving non-monotonic transfer function is correct.

**6. Conclusion and Future Work**

The simulation result shows the novel improvement to LMS proposed in the paper can tackle the effect of MZM raise cosine transfer function, which is symmetric and non-monotonic. The improvement also has generality, and can be extended into wider range of implementations where transfer function is non-monotonic. Current simulation has shown it can achieve a good result when utilized with Lorenz transfer function

The future work contains experiments to further prove the improved method is effective. It will also be applied in some photo-electronic settings, since MZM is a common device in the field.

## 7. Acknowledgement

This paper is developed from the author's Master degree project. The authors would like to acknowledge Professor Yihan Li, who provided further guidance and meticulous revise in paper writing.